\documentclass[12pt]{article}
\usepackage{epsfig}
\usepackage{axodraw4j}

\def\beq{\begin{equation}}
\def\eeq#1{\label{#1}\end{equation}}
\def\eeqn{\end{equation}}

\def\beqa{\begin{eqnarray}}
\def\eeqa#1{\label{#1}\end{eqnarray}}
\def\eeqan{\end{eqnarray}}

\let\bar=\overbar

\def\Dslash{\not{\hbox{\kern-4pt $D$}}}
\def\dslash{\not{\hbox{\kern-2pt $\del$}}}

\def\msb{{\bar{\ssstyle M \kern -1pt S}}}

\def\Title#1{\begin{center} {\Large {\bf #1} } \end{center}}

\begin{document}

\Title{Forward-Backward and Charge Asymmetries in the Standard Model}

\bigskip\bigskip


\begin{raggedright}

{\it Valentin Ahrens\index{Ahrens, V.}\\
Institut f\"ur Physik (THEP)\\
 Johannes Gutenberg-Universit\"at\\ 
55099, Mainz, Germany}
\medskip

{\it Andrea Ferroglia\index{Ferroglia, A.}\\
Physics Department\\
New York City College of Technology\\
11201 Brooklyn, NY, US}
\medskip

{\it Matthias Neubert\index{Neubert, M.}\\
Institut f\"ur Physik (THEP)\\
 Johannes Gutenberg-Universit\"at\\ 
55099, Mainz, Germany}
\medskip

{\it Benjamin Pecjak\index{Pecjak, B.}\\
Institute for Particle Physics Phenomenology\\
University of Durham\\
DH1 3LE Durham, UK}
\medskip

{\it Lilin Yang \index{Yang, L.}\\
Department of Physics and State Key Laboratory of Nuclear Physics and Technology\\
Peking University\\
100871 Beijing, China}
\bigskip
\end{raggedright}


{\abstract{This talk reviews the Standard Model predictions for  the top-quark forward backward and charge asymmetries measured at the Tevatron and at the LHC.
\let\thefootnote\relax\footnote{\emph{Proceedings of CKM 2012, the 7th International Workshop on the CKM Unitarity Triangle, University of Cincinnati, USA, 28 September - 2 October 2012 }}
}}

\section{Top Quark Forward-Backward Asymmetry at the Tevatron}

\begin{figure}[t]
\begin{center}
\begin{tabular}{ccc}
\resizebox{45mm}{!}{
\includegraphics{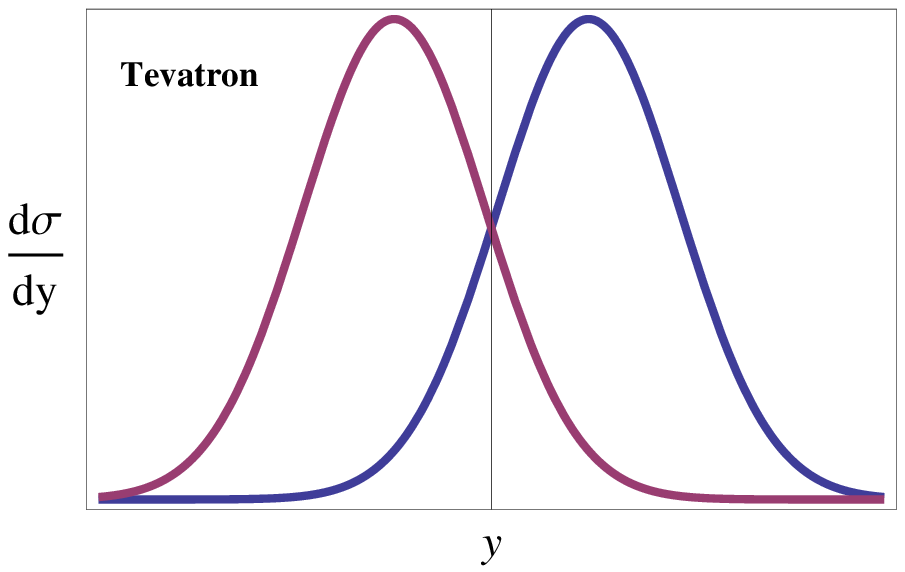}} & \hspace*{1.5cm} & 
\resizebox{45mm}{!}{
\includegraphics{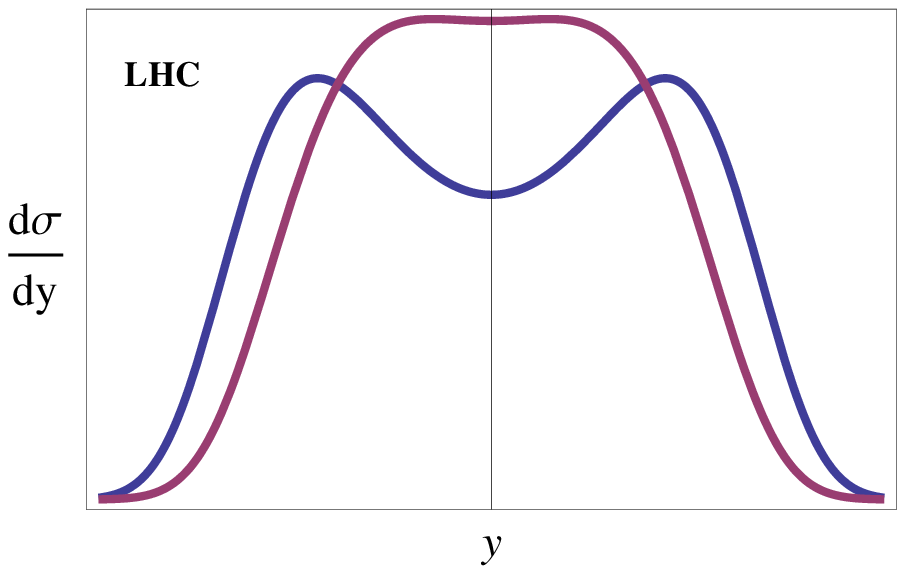}} 
\end{tabular}
\caption{Left Panel:  Sketch of the top and antitop distribution as a function of the rapidity at the Tevatron. 
Right Panel: Sketch of the top and antitop distribution as a function of the rapidity at the LHC. Blue (darker) lines represent the number of top quarks; strawberry (lighter) lines represent the number of antitop quarks.}
\label{fig:TEVcartoon}
\label{fig:LHCcartoon}
\vspace*{-5mm}
\end{center}
\end{figure}

At the Tevatron collider at Fermilab, top quarks were mainly produced in pairs with their antiparticles. 
In proton-antiproton collisions, top quarks are more likely to be produced in the direction of the incoming proton, while antitop quarks are more likely to be produced in the direction of the incoming antiproton (see Fig.~\ref{fig:TEVcartoon}).
The  difference in the production rates of top and antitop quarks in the forward and backward directions can be employed in order to define an  observable asymmetry \cite{Kuhn:1998jr,Kuhn:1998kw}. This top quark forward-backward (FB) asymmetry is defined as
\begin{equation}
A^i_{\mbox{{\tiny FB}}} \equiv \frac{N(y^i_t>  0)-N(y^i_t < 0)}{N(y^i_t > 0)+N(y^i_t < 0)} \, ,
\label{eq:tevfb}
\end{equation}
where $y_t$ is the top quark rapidity. The superscript $i$ indicates the reference frame in which the rapidity is measured;
$i = p \bar{p}$ indicates the laboratory frame, while $i = t \bar{t}$ indicates the top-pair rest frame.
At hadron colliders, the production of top quark pairs is dominated by QCD. The charge conjugation
invariance of the strong interaction implies that the difference in the production of top quarks in the forward and
backward hemispheres is equivalent to the difference in the production of top and antitop quarks in the forward
hemisphere. Therefore, the FB asymmetry at the Tevatron is equivalent to a charge asymmetry defined as
\begin{equation}
A^i \equiv \frac{N(y^i_t> 0)-N(y^i_{\bar{t}} > 0)}{N(y^i_t > 0)+N(y^i_{\bar{t}} > 0)} \, .
\label{eq:tevcharge}
\end{equation}
%
%

In QCD, the first non-vanishing contributions to the numerator of Eqs.~(\ref{eq:tevfb}, \ref{eq:tevcharge}) arises at order $\alpha_s^3$, and the expansion of $A_{\mbox{{\tiny FB}}}$ in powers of $\alpha_s$ gives
\begin{equation}
A_{\mbox{{\tiny FB}}} =  \frac{\alpha_s^3 N_1 + \alpha_s^4 N_2 +\cdots}{\alpha_s^2 D_0 +\alpha_s^3 D_1 +\cdots} = \alpha_s \frac{N_1}{D_0} + \cdots \, . \label{eq:FBexp}
\end{equation}
The origin of $N_1$ is conveniently described by interpreting interferences between Feynman diagrams in terms of cuts of forward scattering amplitudes: The  lowest-order asymmetry numerator $N_1$ arises from diagrams in which 
light-quark fermion lines are connected to the top-quark fermion line by three gluons.  
%
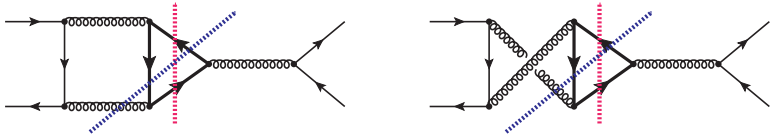
\begin{figure}[tb]
\vspace*{1.5cm}
\begin{center}
  \begin{picture}(0,0) (0,0)
  \SetScale{0.20}
  \SetOffset(-163,-50)
    \SetWidth{3.0}
    \SetColor{Black}
    \Line[arrow,arrowpos=0.5,arrowlength=17.424,arrowwidth=6.97,arrowinset=0.2](96,388)(208,388)
    \Line[arrow,arrowpos=0.5,arrowlength=17.424,arrowwidth=6.97,arrowinset=0.2](208,388)(208,228)
    \Line[arrow,arrowpos=0.5,arrowlength=17.424,arrowwidth=6.97,arrowinset=0.2](208,228)(96,228)
    \SetWidth{6.0}
    \Line[arrow,arrowpos=0.5,arrowlength=27.878,arrowwidth=11.151,arrowinset=0.2](368,388)(368,228)
    \Line[arrow,arrowpos=0.5,arrowlength=27.878,arrowwidth=11.151,arrowinset=0.2,flip](480,308)(368,228)
    \Line[arrow,arrowpos=0.5,arrowlength=27.878,arrowwidth=11.151,arrowinset=0.2,flip](368,388)(480,308)
    \SetWidth{3.0}
    \Gluon(208,388)(368,388){7.5}{11}
    \Gluon(208,228)(368,228){7.5}{11}
    \Gluon(480,308)(640,308){7.5}{11}
    \Line[arrow,arrowpos=0.5,arrowlength=17.424,arrowwidth=6.97,arrowinset=0.2](640,308)(736,388)
    \Line[arrow,arrowpos=0.5,arrowlength=17.424,arrowwidth=6.97,arrowinset=0.2](736,228)(640,308)
        \Line[arrow,arrowpos=0.5,arrowlength=17.424,arrowwidth=6.97,arrowinset=0.2](896,388)(1008,388)
        \Line[arrow,arrowpos=0.5,arrowlength=17.424,arrowwidth=6.97,arrowinset=0.2](1008,388)(1008,228)
        \Line[arrow,arrowpos=0.5,arrowlength=17.424,arrowwidth=6.97,arrowinset=0.2](1008,228)(896,228)
    \SetWidth{6.0}
        \Line[arrow,arrowpos=0.5,arrowlength=27.878,arrowwidth=11.151,arrowinset=0.2](1168,388)(1168,228)
        \Line[arrow,arrowpos=0.5,arrowlength=27.878,arrowwidth=11.151,arrowinset=0.2,flip](1168,388)(1280,308)
        \Line[arrow,arrowpos=0.5,arrowlength=27.878,arrowwidth=11.151,arrowinset=0.2,flip](1280,308)(1168,228)
    \SetWidth{3.0}
\Line[arrow,arrowpos=0.5,arrowlength=17.424,arrowwidth=6.97,arrowinset=0.2](1440,308)(1536,388)
\Line[arrow,arrowpos=0.5,arrowlength=17.424,arrowwidth=6.97,arrowinset=0.2](1536,228)(1440,308)
 \Gluon(1008,388)(1077,320){7.5}{7}
     \Gluon(1008,228)(1168,388){7.5}{16}
         \Gluon(1168,228)(1097,298){7.5}{7}
             \Gluon(1280,308)(1440,308){7.5}{11}
    \SetWidth{1.0}
    \SetColor{Black}
    \Vertex(208,388){6}
    \Vertex(208,228){6}
    \Vertex(368,228){6}
    \Vertex(368,388){6}
    \Vertex(480,308){6}
    \Vertex(640,308){6}
    \Vertex(1008,388){6}
    \Vertex(1008,228){6}
        \Vertex(1168,388){6}
        \Vertex(1168,228){6}
            \Vertex(1440,308){6}
            \Vertex(1280,308){6}

        \SetWidth{8.0}
        \SetColor{WildStrawberry}
        \Line[dash,dashsize=4](416,420)(416,196)
                \Line[dash,dashsize=4](1216,420)(1216,196)
    \SetWidth{8.0}
    \SetColor{Blue}
    \Line[dash,dashsize=4](528,404)(256,180)
        \Line[dash,dashsize=4](1312,404)(1040,180)
  \end{picture}
  \vspace*{2mm}
\caption{Quark annihilation diagrams contributing to the asymmetry numerator at lowest order in QCD.
Thin arrow lines indicate light quarks, thick arrow lines indicate top quarks.
The vertical dashed cuts indicate the interferences between a one-loop box and the tree level diagram. The diagonal dashed lines split the forward amplitudes in two $2 \to 3$ tree level diagrams.}
\label{fig:LOQCD}
\vspace*{-4mm}
\end{center}
\end{figure}
%
Fig.~\ref{fig:LOQCD} shows that, 
in the quark annihilation channel, this happens in the interference between one-loop box diagrams and the single tree-level
diagram and in the interference of two $2 \to 3$ tree level diagrams. These contributions to $N_1$ are proportional to the square of the totally symmetric color coefficient $d_{abc}$\cite{Kuhn:1998jr,Kuhn:1998kw}.
In addition to the numerically dominant quark annihilation diagrams,  another,  numerically negligible contribution to $N_1$ arises from the flavor excitation channel $gq \to t \bar{t} q$.
The gluon fusion channel does not contribute to the numerator of the FB asymmetry at any order in perturbation theory, since gluon fusion is charge symmetric. At lowest order in perturbation theory,\footnote{In this talk, we label the various contributions to the asymmetry with reference to the order in $\alpha_s$ at which the numerator in Eq.~(\ref{eq:FBexp}) is evaluated, in comparison to $\alpha_s^2$. Therefore, the first non-vanishing asymmetry is referred to here as NLO asymmetry.}
the QCD asymmetry is \cite{Kuhn:2011ri}
\begin{equation}
\left[A_{\mbox{{\tiny FB}}}^{p\bar{p}}\right]_{\mbox{{\footnotesize QCD}}}^{\mbox{{\footnotesize NLO}}} = (4.7 \pm 0.7) \%  \, , \qquad
\left[A_{ \mbox{{\tiny FB}}}^{t\bar{t}}\right]_{ \mbox{{\footnotesize QCD}}}^{\mbox{{\footnotesize NLO}}}  = (7.2 \pm 0.9) \% \, .
\end{equation}
The predictions for the asymmetry at NLO in QCD have been consistently lower than the values measured at CDF and D0 for several years. The tension between theory and experiment is of the order of two standard deviations \cite{JWilson}.
The situation became even more intriguing in January 2011, when CDF reported separate asymmetry measurements in the 
$t \bar{t}$ rest frame for events with a top-pair invariant mass $M_{t\bar{t}}$ smaller or larger than $450$~GeV \cite{Aaltonen:2011kc}. In the high invariant mass bin, the measured asymmetry was more than three standard deviations larger than the NLO QCD  prediction \cite{Kuhn:2011ri}
\begin{equation}
\left[A_{\mbox{{\tiny FB}}}^{t \bar{t}}\right]_{M_{t\bar{t}} \ge 450~\mbox{{\tiny GeV}}}^{\mbox{{\footnotesize NLO QCD}}} = 10.6 \pm 1.1 \%
\, .
\end{equation}
A more recent and more precise measurement of the same quantity at CDF \cite{CDFnote},
$
\left[A_{\mbox{{\tiny FB}}}^{t \bar{t}} \right]_{M_{t\bar{t}} \ge 450~\mbox{{\tiny GeV}}}^{\mbox{{\footnotesize exp.}}} = 29.6 \pm 6.7   \%
\, ,
$
gives rise to a slightly smaller discrepancy of $\sim 2.8 \,\sigma$. For both the total and high pair invariant mass asymmetry, explanations of the discrepancy in terms of  Physics Beyond the Standard Model (SM) should not spoil the good agreement between theory and experiment in other top quark related observables, such as the total pair production cross section \cite{CDelaunay}. Of course, it is still important to ensure that large higher order radiative corrections in the SM are not overlooked. Several groups investigated this aspect in the last few years.

\section{SM Predictions for the FB Asymmetry}

The complete mixed QED/QCD and electroweak (EW)/QCD corrections of order $\alpha^2_s \alpha$ were evaluated by Hollik and Pagani \cite{Hollik:2011ps}, and subsequently by K\"uhn and Rodrigo \cite{Kuhn:2011ri}. These corrections arise from diagrams 
similar to the ones in Fig.~\ref{fig:LOQCD}, in which one of the gluon lines is replaced by a photon or a $Z$ boson. 
%
\begin{table}[bt]
\begin{center}
\begin{tabular}{l|c|c|c}  
Corrections &  $A_{\mbox{{\tiny FB}}}^{p \bar{p}}$  &  $A_{\mbox{{\tiny FB}}}^{t \bar{t}}$ &
$A_{\mbox{{\tiny FB}}}^{t \bar{t}}$ for  ${M_{t\bar{t}} \ge 450~\mbox{GeV}}$
 \\ \hline
NLO QCD  &   $4.7 \pm 0.7 \%$     &     $7.2 \pm 0.9 \%$      &     $10.6 \pm 1.1 \%$  \\
QCD-QED $u \bar{u}$ &   $0.94 \%$    &   $1.45 \%$     &  $2.19 \%$ \\ 
QCD-QED $d \bar{d}$ & $-0.08 \%$    &   $-0.12 \%$    &  $-0.15 \%$   \\ 
QCD-EW $u \bar{u}$ &   $0.11 \%$    &  $0.18 \%$  &  $0.27 \%$  \\
QCD-EW $d \bar{d}$ &   $-0.03 \%$   & $-0.05 \%$  &  $-0.06 \%$  \\ \hline
Total SM &  $5.6 \pm 0.7 \%$     & $8.7 \pm 1 \%$    & $12.8 \pm 1.1 \%$   \\ \hline
\end{tabular}
\caption{NLO QCD and mixed QED/QCD  and EW/QCD contributions to the asymmetry in the $p \bar{p}$ frame, in the $t\bar{t}$ frame, and in the high pair invariant mass bin. Contributions at order $\alpha_s^2 \alpha$ are listed separately for incoming $u$ and $d$ quarks. The last line includes the sum of all of the  contributions. The numbers are taken  from \cite{Kuhn:2011ri}, where MSTW2008 PDFs were employed.}
\label{tab:QEDEW}
\vspace*{-6mm}
\end{center}
\end{table}
%
Table~\ref{tab:QEDEW} shows that the  dominant numerical effect is given by photon corrections, and that the total effect of the $\alpha^2_s \alpha$ corrections increases the NLO QCD asymmetry by a factor $\sim 1.2$, irrespectively from the reference frame, and both for the total and high pair invariant mass asymmetry.

It is also important to assess the impact of higher order QCD corrections, especially given the fact that only the lowest order contribution to the asymmetry in fixed order perturbation theory is currently known. 
In light of the recent calculation of the NNLO corrections to the total top-quark pair-production cross section in the quark-annihilation  and quark-gluon channels \cite{Baernreuther:2012ws, Czakon:2012zr, Czakon:2012pz},
an evaluation of the coefficient $N_2$ in (\ref{eq:FBexp}) and of the NNLO asymmetry should be possible in the near future. It must be observed that a calculation of the asymmetry at NNLO does not require an evaluation of the NNLO corrections to the gluon fusion channel.

At the moment, a full evaluation of the asymmetry at NNLO is not yet available. Several top-pair differential distributions have been evaluated in renormalization group improved perturbation theory up to NNLL accuracy.
In particular, the NNLL resummation of soft gluon emission effects was carried out for the pair invariant mass distribution 
\cite{Ahrens:2010zv} and for the top quark transverse momentum and rapidity distributions \cite{Ahrens:2011mw, Kidonakis:2011zn}.
By integrating these differential distributions, it is possible to obtain predictions for the top-quark FB asymmetry at NNLL accuracy \cite{Ahrens:2011uf}. (A study of the FB asymmetry at NLL accuracy was carried out in \cite{Almeida:2008ug}.) In particular, by starting from the top quark $p_T$ and rapidity distributions, one can evaluate the asymmetry in the laboratory frame. Instead, by starting from the pair invariant mass distribution, one can obtain the asymmetry in the $t \bar{t}$ rest frame.

The resummation of soft gluon emission corrections is based upon the fact that in partonic events near the kinematic threshold, the partonic cross section receives large logarithmic corrections which depend upon a variable which parameterizes
the distance from the threshold. In Pair Invariant Mass (PIM) kinematics, which is employed to study the pair invariant mass distribution, the parameter of interest is $z = M_{t \bar{t}}^2 /\hat{s}$, where $M_{t \bar{t}}$ is the pair invariant mass while $\hat{s}$ is 
the partonic center of mass energy. In the PIM threshold limit, one has that $z \to 1$. In order to study the top quark $p_T$ and rapidity distribution, one employs One Particle Inclusive (1PI) kinematics, where the variable 
parameterizing the distance from threshold is $s_4 = (p_4 + k)^2 -m_t^2$. $p_4$ and $k$ are the four momenta of the 
antitop  and of the gluon radiation, respectively. Near threshold, $s_4 \to 0$. In the soft gluon limit, the partonic cross section factors into the 
product of a hard function (which depends only on virtual correction) and a soft function (which depends only on soft gluon 
emission). While the hard function is the same in both PIM and 1PI kinematics, the soft functions in PIM and 1PI kinematics are 
different. The large logarithmic corrections associated with soft gluon emission can be resummed by solving the 
renormalization group equations satisfied by the soft and hard functions directly in momentum space 
\cite{Ahrens:2010zv, Ahrens:2011mw}. In spite of the fact that  one needs to integrate over a partonic phase space which is larger than the threshold region in order to obtain hadronic observables, the rapid fall-off of the partonic luminosity away from the threshold makes the soft gluon emission corrections numerically dominant at the level of hadronic observables.
This mechanism goes under the name of dynamical threshold enhancement.
%
\begin{table}[bt]
\begin{center}
\begin{tabular}{l|c|c|}  
MSTW2008 & $\Delta\sigma^{p\bar{p}}_{\mbox{{\tiny FB}}}$ [pb]  &   $A_{\mbox{{\tiny FB}}}^{p \bar{p}}$ [$\%$] 
 \\ \hline
 NLO QCD & 0.260{\footnotesize $^{+0.141+0.020}_{-0.084-0.014}$} & 4.81{\footnotesize $^{+0.45+0.13}_{-0.39-0.13}$} \\
 NLO+NNLL & 0.312{\footnotesize $^{+0.027+0.023}_{-0.035-0.019}$} & 4.88{\footnotesize $^{+0.20+0.17}_{-0.23-0.18}$}\\
\end{tabular}
\caption{The asymmetric cross section and FB asymmetry in the $p\bar{p}$ frame. The first error refers 
to perturbative uncertainties, while the second error indicates the PDF uncertainty.  Results for other PDF choices are similar and can be found in \cite{Ahrens:2011uf}.}
\label{tab:ppAFB}
\vspace*{-6mm}
\end{center}
\end{table}
%

In the laboratory frame, one can define an asymmetric cross section as 
\begin{equation}
\label{eq:fbasympp}
\Delta\sigma^{p\bar{p}}_{\mbox{{\tiny FB}}} \equiv \int_0^{y^+_t} dy_t 
\left[
 \int_0^{p_T^{\mbox{{\tiny max}}}} dp_T \, \frac{d^2 \sigma^{p\bar{p} \to t
X_{\bar{t}}}}{dp_Tdy_t} - \int_0^{p_T^{\mbox{{\tiny max}}}} dp_T \,\frac{d^2
\sigma^{p\bar{p} \to t X_{\bar{t}}}}{dp_Td\bar{y}_t} \Bigg|_{\bar{y}_t=-y_t} \right] \, ,
\end{equation}
where the extrema of the integration region are
\begin{equation}
\label{eq:yplimits}
  y^+_t = \frac{1}{2} \ln\frac{1+\sqrt{1-4m_t^2/s}}{1-\sqrt{1-4m_t^2/s}} \quad \mbox{and}
  \quad p_T^{\mbox{{\tiny max}}} = \frac{\sqrt{s}}{2}\sqrt{ \frac{1}{\cosh^2 y_t}-\frac{4m_t^2}{s}} \, .
\end{equation}
Here, $s$ indicates the square of the hadronic center of mass energy and $y_t$ indicates the top quark rapidity.
The total asymmetry is then obtained by taking the ratio of the asymmetric cross section and the total cross section, appropriately expanded in $\alpha_s$. The effect of the NNLL corrections on the asymmetric cross section in Eq.~(\ref{eq:fbasympp}) and on the total asymmetry can be found in Table~\ref{tab:ppAFB}. The results shown refer to the use of the
MSTW2008 PDF set. The central values are
obtained by fixing the factorization scale at $\mu_f = m_t$, and the scale uncertainties are estimated by varying $\mu_f$ between
$m_t/2$ and $2 m_t$. In NNLL calculations the hard and soft scales are also varied by following the procedure
adopted in \cite{Ahrens:2011mw}. The PDF uncertainties were estimated by iterating through the $90 \%$
confidence level (CL) sets. One can observe that:  \emph{i)} The PDF uncertainties for
the asymmetry, expressed as
a percentage of the central values, are about half as large as those for the asymmetric
cross section. \emph{ii)} NNLL accuracy results for
the asymmetric cross section
are numerically consistent
with the NLO results for
$\mu_f = m_t$, while the scale
uncertainty is reduced by
more than a factor of 2.  \emph{iii)} The NLO+NNLL central value for the
asymmetry does not change
significantly with respect to
the NLO predictions, and the
scale uncertainties are
reduced.

Starting from the pair invariant mass distribution, one can define an asymmetric cross section in the $t \bar{t}$ rest frame as follows:
\begin{equation}
  \Delta\sigma^{t\bar{t}}_{\mbox{{\tiny FB}}} \equiv \int_{2m_t}^{\sqrt{s}} dM_{t\bar{t}} \left[
    \int_0^1 d\cos\theta \, \frac{d^2\sigma^{p\bar{p} \to t\bar{t}
        X}}{dM_{t\bar{t}}d\cos\theta} - \int_{-1}^0 d\cos\theta \,
    \frac{d^2\sigma^{p\bar{p} \to t\bar{t} X}}{dM_{t\bar{t}}d\cos\theta} \right] \, .
    \label{eq:ttacs}
\end{equation}
Also in this case, the total asymmetry is obtained by dividing the asymmetric cross section by the total cross section.
The effect of the NNLL corrections can be seen in Table~\ref{tab:ttAFB}, where MST2008 PDFs are employed. 
%
\begin{table}[tb]
\begin{center}
\begin{tabular}{l|c|c|c|c}  
MSTW2008 & $\Delta\sigma^{t\bar{t}}_{\mbox{{\tiny FB}}}$ [pb]  &   $A_{\mbox{{\tiny FB}}}^{t \bar{t}}$  [$\%$]
 &$[A_{\mbox{{\tiny FB}}}^{t \bar{t}}]_{M_{t \bar{t}} < 450~\mbox{{\tiny GeV}}}$ &$[A_{\mbox{{\tiny FB}}}^{t \bar{t}}]_{M_{t \bar{t}} > 450~\mbox{{\tiny GeV}}}$  \\ \hline
 NLO QCD &0.395{\footnotesize $^{+0.213+0.028}_{-0.128-0.021}$}  & 7.32{\footnotesize $^{+0.69+0.18}_{-0.59-0.19}$} &  5.2{\footnotesize $^{+0.6}_{-0.2}$} & 10.8{\footnotesize $^{+1.0}_{-0.8}$}\\
 NLO+NNLL & 0.448{\footnotesize $^{+0.080+0.030}_{-0.071-0.026}$}  & 7.24{\footnotesize $^{+1.04+0.20}_{-0.67-0.27}$} & 5.2{\footnotesize $^{+0.9}_{-0.6}$} &11.1{\footnotesize $^{+1.7}_{-0.9}$} \\
\end{tabular}
\caption{The asymmetric cross section and FB asymmetry (total and binned with respect to $M_{t \bar{t}}$) in the $t\bar{t}$ frame. The first error refers 
to perturbative uncertainties. The second error (if present) is the PDF uncertainty.}
\label{tab:ttAFB}
\vspace*{-6mm}
\end{center}
\end{table}
%
One sees that the scale uncertainties in
the asymmetric cross
section are roughly halved
at NLO+NNLL order
compared to NLO. The scale uncertainties on
the FB asymmetry increase
slightly after the NNLL
resummation, while the
central values are nearly
unchanged; therefore, one should be cautious of the rather small scale
uncertainties in the NLO calculation of the asymmetry,
which result from large cancellations
in the ratio not observed in the resummed result.
Prediction for the asymmetry for events falling in a specific pair invariant mass bin can be obtained by changing  the $M_{t \bar{t}}$ integration interval in Eq.~(\ref{eq:ttacs})  (as well as in  the asymmetry's denominator) from $[2 m_t, \sqrt{s}]$
to $[m_1, m_2]$, where $m_1$ and $m_2$ are the extrema of the desired invariant mass bin. Results for the asymmetry in the 
bins $M_{t \bar{t}} < 450$~GeV and $M_{t \bar{t}} > 450$~GeV can also be found in Table \ref{tab:ttAFB}. In both bins, the FB asymmetries are essentially
unchanged by NNLL resummation. Results obtained by employing other PDF sets can be found in \cite{Ahrens:2011uf} and are very consistent with the ones shown in Table \ref{tab:ttAFB}. It is therefore possible to conclude that neither soft-gluon resummation effects, nor systematic PDF uncertainties
reduce the discrepancy between theory and experiment in the high invariant-mass bin. In the low invariant mass bin, the measured asymmetry  $
\left[A_{\mbox{{\tiny FB}}}^{t \bar{t}} \right]_{M_{t\bar{t}} < 450~\mbox{{\tiny GeV}}}^{\mbox{{\footnotesize exp.}}} = 11.6 \pm 15.3   \%$ \cite{CDFnote} is about one standard deviation lower than the predicted asymmetry.

\section{Top Charge Asymmetry at the LHC}

Since the initial state is symmetric at the LHC, the asymmetry defined as in Eq.~(\ref{eq:tevfb}) vanishes. However, at the partonic level, top quarks are preferably emitted in the  direction of the incoming quark, while antitops are preferably emitted in the direction of the incoming antiquark. Since valence quarks carry more momentum than sea antiquarks, top quarks tend to be emitted at larger rapidities than antiquarks. The situation is sketched in Fig.~\ref{fig:LHCcartoon} (right panel). This feature can be employed to define a charge asymmetry at the LHC:
\begin{equation}
A_C^y = \frac{N\left( \Delta |y| > 0\right) - N\left( \Delta |y| < 0\right)}{N\left( \Delta |y| > 0\right) + N\left( \Delta |y| < 0\right)} \, , \label{eq:chlhc}
\end{equation}
with $\Delta |y|$ defined as the difference between the absolute value of the top and antitop rapidities $\Delta |y|  = |y_t| - |y_{\bar{t}}|$.
%
\begin{table}[tb]
\begin{center}
\begin{tabular}{c|c||c|c|c}  
  7 TeV ATLAS & 7 TeV CMS & SM 7 TeV & SM 8 TeV & SM 14 TeV \\ \hline
  $0.4 \pm 1.0 \pm 1.2 \% $ & $-1.9 \pm 2.8 \pm 2.4 \%$ & $1.15 \pm 0.06 \%$ & $1.02 \pm 0.05 \%$ & $0.59 \pm 0.03 \%$
\end{tabular}
\caption{Measured charge asymmetry at the LHC with $\sqrt{s} =  7$~TeV; the first error is statistical, while the second error is systematic. Predicted charge asymmetry at the LHC in the SM for different center of mass energies \cite{Kuhn:2011ri, Rodrigo:2012as}.}
\label{tab:LHC}
\vspace*{-6mm}
\end{center}
\end{table}
%
The observable in Eq.~(\ref{eq:chlhc}) was measured both by ATLAS and CMS \cite{ATLAS:2012an, CMS} for a center of mass energy of $7$~TeV. Table~\ref{tab:LHC} shows the measurement results together with the SM predictions for several values of $\sqrt{s}$ \cite{Kuhn:2011ri, Rodrigo:2012as}. Both experiments obtain 
asymmetries which are compatible with, but smaller than, the SM prediction. Because of the large number of top-pair events at the LHC, the measurements of the asymmetry should soon be dominated by systematics. 
As the LHC center of mass energy increases, the asymmetry decreases; this is because the top pair production  becomes increasingly dominated by the symmetric gluon fusion channel. The asymmetry can be enhanced by imposing cuts to select events with large rapidities or large $M_{t \bar{t}}$. 
Since the calculation of the charge asymmetry requires  information regarding the rapidity of both the top quark and the antitop quark, the results in \cite{Ahrens:2011mw} cannot be employed to calculate the LHC charge asymmetry at NNLL accuracy.

\bigskip
The work of A. Ferroglia was supported in part by the PSC-CUNY Award N. 65214-00-43 and by National Science Foundation Grant No. PHY-1068317.


\begin{thebibliography}{99}


\bibitem{Kuhn:1998jr} 
  J.~H.~Kuhn and G.~Rodrigo,
  Phys.\ Rev.\ Lett.\  {\bf 81}, 49 (1998)

\bibitem{Kuhn:1998kw} 
  J.~H.~Kuhn and G.~Rodrigo,
  Phys.\ Rev.\ D {\bf 59}, 054017 (1999)

\bibitem{Kuhn:2011ri} 
  J.~H.~Kuhn and G.~Rodrigo,
  JHEP {\bf 1201}, 063 (2012)

\bibitem{JWilson}
J. Wilson, these proceedings.

\bibitem{Aaltonen:2011kc} 
  T.~Aaltonen {\it et al.}  [CDF Collaboration],
  Phys.\ Rev.\ D {\bf 83}, 112003 (2011)
  
  \bibitem{CDFnote}
CDF Collaboration, CDF Note 10807, (2012)
  
  
  \bibitem{CDelaunay}
  C. Delaunay, these proceedings.
 
 \bibitem{Hollik:2011ps} 
   W.~Hollik and D.~Pagani,
   Phys.\ Rev.\ D {\bf 84}, 093003 (2011)

 \bibitem{Baernreuther:2012ws} 
   P.~Baernreuther, M.~Czakon and A.~Mitov,
   Phys.\ Rev.\ Lett.\  {\bf 109}, 132001 (2012)
 
 \bibitem{Czakon:2012zr} 
   M.~Czakon, A.~Mitov and A.~Mitov,
   arXiv:1207.0236 [hep-ph].
 
 \bibitem{Czakon:2012pz} 
   M.~Czakon and A.~Mitov,
   arXiv:1210.6832 [hep-ph].
 
 \bibitem{Ahrens:2010zv} 
   V.~Ahrens \emph{et al.}
   JHEP {\bf 1009}, 097 (2010)
 
 \bibitem{Ahrens:2011mw} 
   V.~Ahrens \emph{et al.}
   JHEP {\bf 1109}, 070 (2011)
   
   \bibitem{Kidonakis:2011zn} 
     N.~Kidonakis,
     Phys.\ Rev.\ D {\bf 84}, 011504 (2011)
   
   \bibitem{Ahrens:2011uf} 
     V.~Ahrens \emph{et al.}
     Phys.\ Rev.\ D {\bf 84}, 074004 (2011)
     
     \bibitem{Almeida:2008ug} 
       L.~G.~Almeida, G.~F.~Sterman and W.~Vogelsang,
       Phys.\ Rev.\ D {\bf 78}, 014008 (2008)

       
       
       \bibitem{ATLAS:2012an} 
         G.~Aad {\it et al.}  [ATLAS Collaboration],
         Eur.\ Phys.\ J.\ C {\bf 72}, 2039 (2012)
 
 \bibitem{CMS}
 [CMS Collaboration], CMS-PAS-TOP-11-030.
 
 
        \bibitem{Rodrigo:2012as} 
          G.~Rodrigo,
          arXiv:1207.0331 [hep-ph].
 
 
\end{thebibliography}
\end{document}